\documentclass{PoS}
\newcommand{ \psirp }{\Psi_{RP}}
\newcommand{ \phia }{\phi_{\alpha}}
\newcommand{ \phib }{\phi_{\beta}}
\newcommand{ \mean }[1]{\left\langle #1 \right\rangle}

\newcommand{\SPD}{\rm{SPD}}

\newcommand{\TPC}{\rm{TPC}}

\newcommand{\VZERO}{\rm{VZERO}}
\newcommand{\ZDC}{\rm{ZDC}}

\newcounter{vers}

\title{ALICE probes of local parity violation with charge dependent azimuthal correlations in Pb-Pb collisions}

\ShortTitle{Charge dependent azimuthal correlations in Pb-Pb collisions}

\author{\speaker{Ilya Selyuzhenkov}  for the ALICE Collaboration\\
        Research Division and ExtreMe Matter Institute EMMI,\\
GSI Helmholtzzentrum f\"ur Schwerionenforschung, Darmstadt, Germany
        E-mail: \email{ilya.selyuzhenkov@gmail.com}}

\abstract{

We report on the measurement of two and three particle azimuthal correlations
in Pb-Pb collisions at $\sqrt{s_{NN}}$ = 2.76 TeV recorded with ALICE at the LHC.
While two particle azimuthal correlations mainly provide
an important information on possible background correlations,
the three particle correlator probes the charge separation of hadrons
with respect to the collision reaction plane which
is expected for local parity violation in strong interactions.
The two and three particle correlations are presented
as a function of collision centrality and differentially vs.
pseudorapidity and transverse momentum,
and provide strong constraints on the possible mechanism of background
(parity conserving) and signal (parity odd) effects in heavy-ion collisions.
}

\FullConference{The Seventh Workshop on Particle Correlations and Femtoscopy\\
		 September 20 - 24 2011\\
		 University of Tokyo, Japan}

\begin{document}

\section{\label{Section:Introduction}Introduction}

An extreme magnetic field created during a non-central relativistic heavy-ion collision
may spontaneously excite instantons and sphalerons from the QCD vacuum
which violates parity symmetry of the strong interactions.
It is argued by Kharzeev {\it et al.} \cite{Ref:Kharzeev} that this may
result in the experimentally observable separation of charges along the magnetic field.
Thus, a measurement of charge separation with respect to the reaction plane for
hadrons produced
in heavy-ion collisions provides a unique way to experimentally probe
the phenomena of local parity violation in strong interactions which
has been discussed for many years \cite{Ref:Kharzeev,Ref:TDLee,Ref:Morley}.
The measurement of the charge dependent correlations by the STAR 
Collaboration~\cite{Ref:STAR} revealed a signal which is qualitatively consistent
with the expectation for the charge separation from local parity violation.
This triggered an intensive discussion about possible interpretation of the
correlations \cite{Ref:Muller,Ref:Pratt,Ref:Koch,Ref:Zhitnitsky}.
In these proceedings I report on the charge dependent 
azimuthal correlations measured at mid-rapidity
for Pb-Pb collisions at $\sqrt{s_{NN}} = 2.76$~TeV with the ALICE 
detector \cite{Ref:ALICE, Ref:ALICEPPR} at the LHC.

\section{\label{Section:Analysis}Observables and experimental data}
An experimental observable which is sensitive to effects from local parity violation
in strong interactions was proposed in \cite{Ref:Sergey3particleCorrelator}:
\begin{eqnarray}
\langle \cos(\phi_{\alpha} + \phi_{\beta} - 2\Psi_{RP}) \rangle =
\mean{\cos\Delta \phia\, \cos\Delta \phib}-\mean{\sin\Delta \phia\,\sin\Delta \phib}.
\label{Eq:3ParticleCorrelator}
\end{eqnarray}
Here $\phi_{\alpha,\beta}$ are the positive or negative charged particle azimuthal angles,
$\Delta \phi=\phi-\psirp$, and $\psirp$ is the reaction plane angle.
Experimentally the reaction plane is estimated
from the azimuthal asymmetry of the produced particles
and~(\ref{Eq:3ParticleCorrelator}) become a three particle correlator.
In addition to~(\ref{Eq:3ParticleCorrelator}) , the measurement of the two-particle correlator
\begin{eqnarray}
\langle \cos(\phi_{\alpha} - \phi_{\beta} \rangle =
\mean{\cos\Delta \phia\, \cos\Delta \phib}+\mean{\sin\Delta \phia\,\sin\Delta \phib}
\label{Eq:2ParticleCorrelator}
\end{eqnarray}
provides an experimental constraint on models
of possible parity-even background contributions to~(\ref{Eq:3ParticleCorrelator}).
Together, the two (\ref{Eq:2ParticleCorrelator}) and three
(\ref{Eq:3ParticleCorrelator}) particle correlators supply information
about in-plane, $\mean{\cos\Delta \phia\, \cos\Delta \phib}$,
and out-of-plane, $\mean{\sin\Delta \phia\,\sin\Delta \phib}$, charge correlations.

The ALICE detector description and performance is given in \cite{Ref:ALICE}.
For this study the Time Projection Chamber (\TPC)
\cite{Ref:ALICETPC}, two forward scintillator arrays (\VZERO) with
pseudorapidity coverage $-3.7 < \eta < -1.7$ and $2.8 < \eta < 5.1$, and
two neutron Zero Degree Calorimeters (\ZDC) \cite{Ref:ALICE}
located at about 114~m on both sides of the interaction point are used.
The TPC, \VZERO~detectors, and neutron \ZDC{s}
are used to estimate the orientation of the reaction plane.
We used a sample of about 13~million minimum-bias Pb-Pb
collisions at $\sqrt{s_{NN}} = 2.76$~TeV recorded by the ALICE detector in 2010.
The standard \mbox{ALICE} event selection criteria with
the collision centrality estimated from \VZERO~detectors
\cite{Ref:AliceFlow} and a collision vertex cut of 7~cm along the beam axis were applied.
Only charged particles with pseudorapidity $|\eta| < 0.8$ and transverse momentum $p_{\rm{T}} >
0.2$~GeV/$c$ are used for the measurements.

The systematic uncertainties were evaluated based on the
comparison between results obtained
with opposite magnetic field polarities and
variation of the results with the cut on the collision vertex along the beam direction within $\pm$10~cm.
The bias in the collision centrality determination is estimated from the difference
between results with the centrality estimated from \VZERO{s}, \TPC~and Silicon Pixel Detector (\SPD).
Contamination from charged tracks which did not originate from the collision vertex was estimated
by varying the cut on the $dca$ within 4~cm.
Effects due to non-uniform acceptance of the \TPC~are
found to be below 2\%~and are corrected for in the final result.
The total systematic uncertainty is obtained by adding uncertainties from all the sources in quadrature.

\section{\label{Section:Results}Results}
Figure~\ref{fig:results}(a) shows the 3-particle
correlator~(\ref{Eq:3ParticleCorrelator}) as a function of centrality
calculated from four different analyses:
\begin{figure}[ht]
\includegraphics[width=0.5\linewidth]{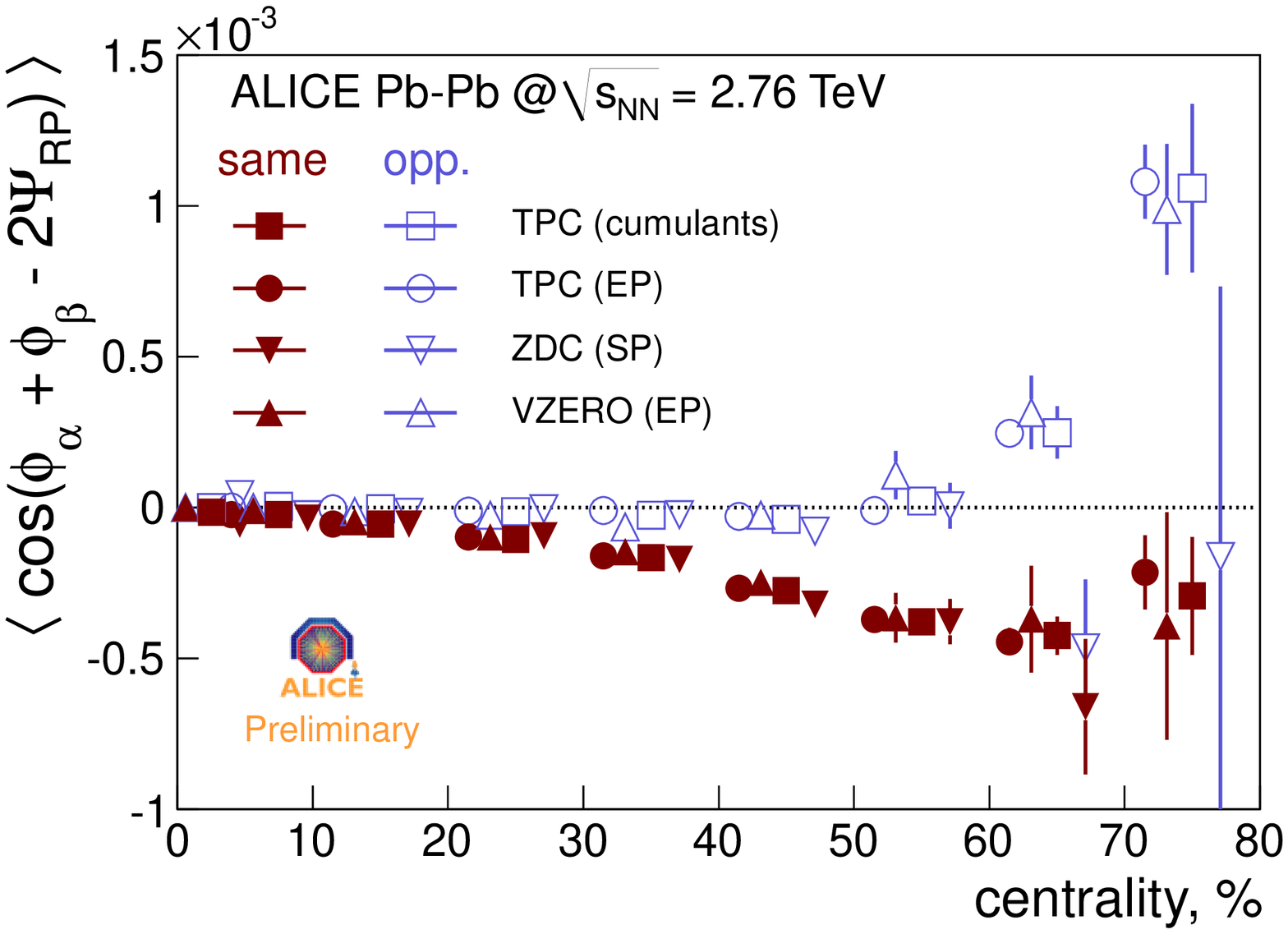}%
\includegraphics[width=0.5\linewidth]{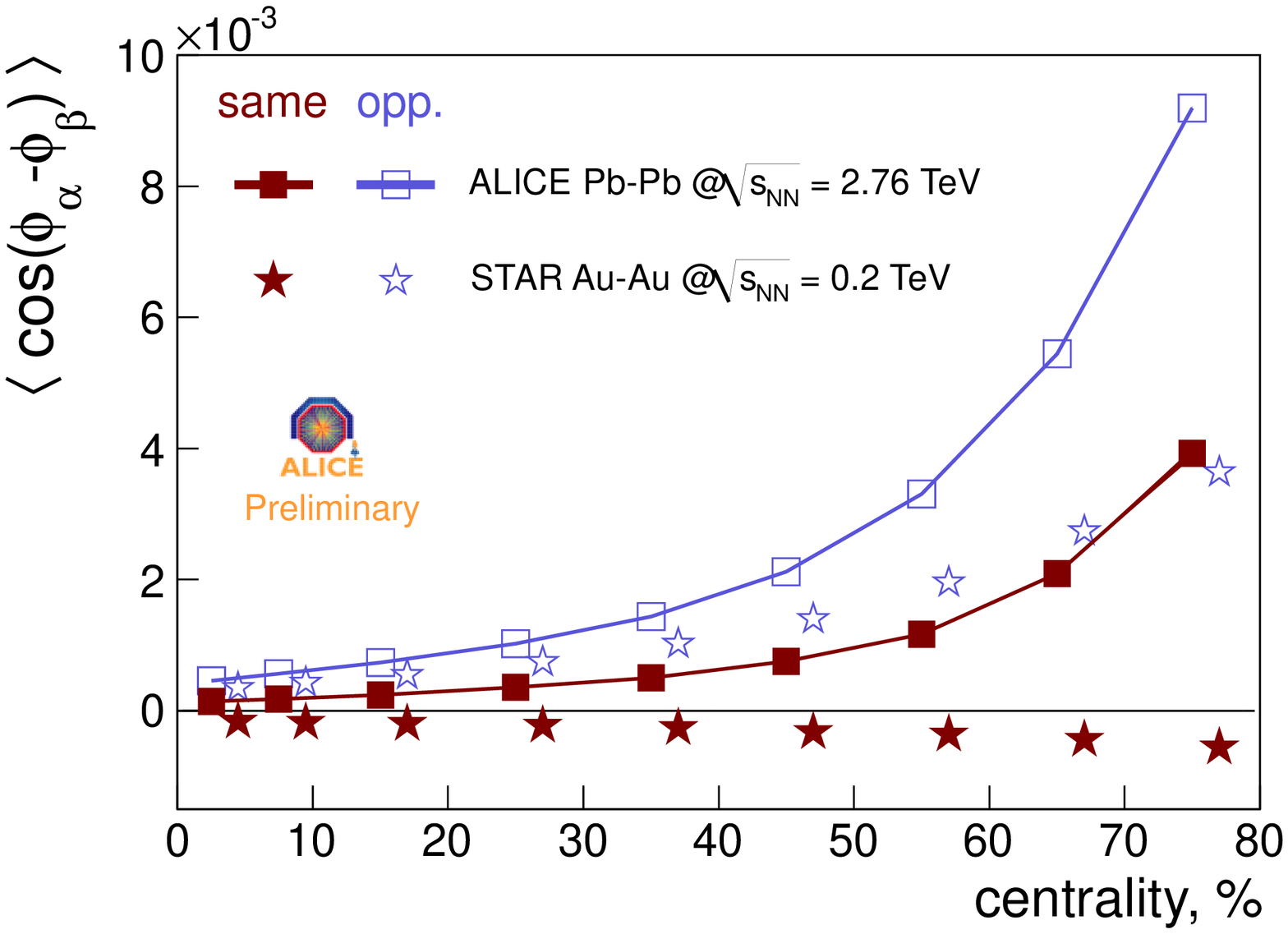}
{\mbox{}\vspace{-0.4cm}
\hspace{3.7cm}\mbox{~} \bf (a)
\hspace{+6.8cm}\mbox{~} \bf (b)}
{\mbox{}\\\vspace{-0.1cm}}
\caption{
(Color online) (a) Centrality dependence of the 3-particle correlator
measured with the reaction plane estimated from \TPC, \ZDC{s}~and the \VZERO{s}~detectors.
(b) Centrality dependence of the 2-particle
correlations compared with the STAR data~\cite{Ref:STAR}.
}
\label{fig:methodComparison}
\label{fig:results}
\end{figure}
3 particle cumulants~\cite{Ref:Qumulants} and a set of three independent measurements with the reaction
plane estimated from different ALICE detector subsystems (\TPC, \ZDC{s} and \VZERO{s})
using the scalar product (SP) and the event plane (EP) measurement techniques \cite{Voloshin:2008dg}.
Agreement between four methods indicate that the measured three particle correlations
represent two charged particle correlation with respect to the reaction plane.
Small variations between methods were used to evaluate the total systematic uncertainty.
The results for positive-positive and negative-negative pairs
were combined into the same charge correlations.
The observed difference in Fig.~\ref{fig:results}(a) between the same and opposite charge correlations
may in part originates from effects of local parity violation in the strong interaction.
However, a number of other charge dependent effects which preserve parity symmetry can also contribute.
To constrain the possible backgrounds experimentally, in Fig.~\ref{fig:results}(b)
we present the centrality dependence of the 2-particle
correlator~(\ref{Eq:2ParticleCorrelator}).
The ALICE results for the 2-particle correlations differ from
those reported by the STAR Collaboration~\cite{Ref:STAR}.

\begin{figure}[ht]
\includegraphics[width=0.5\linewidth]{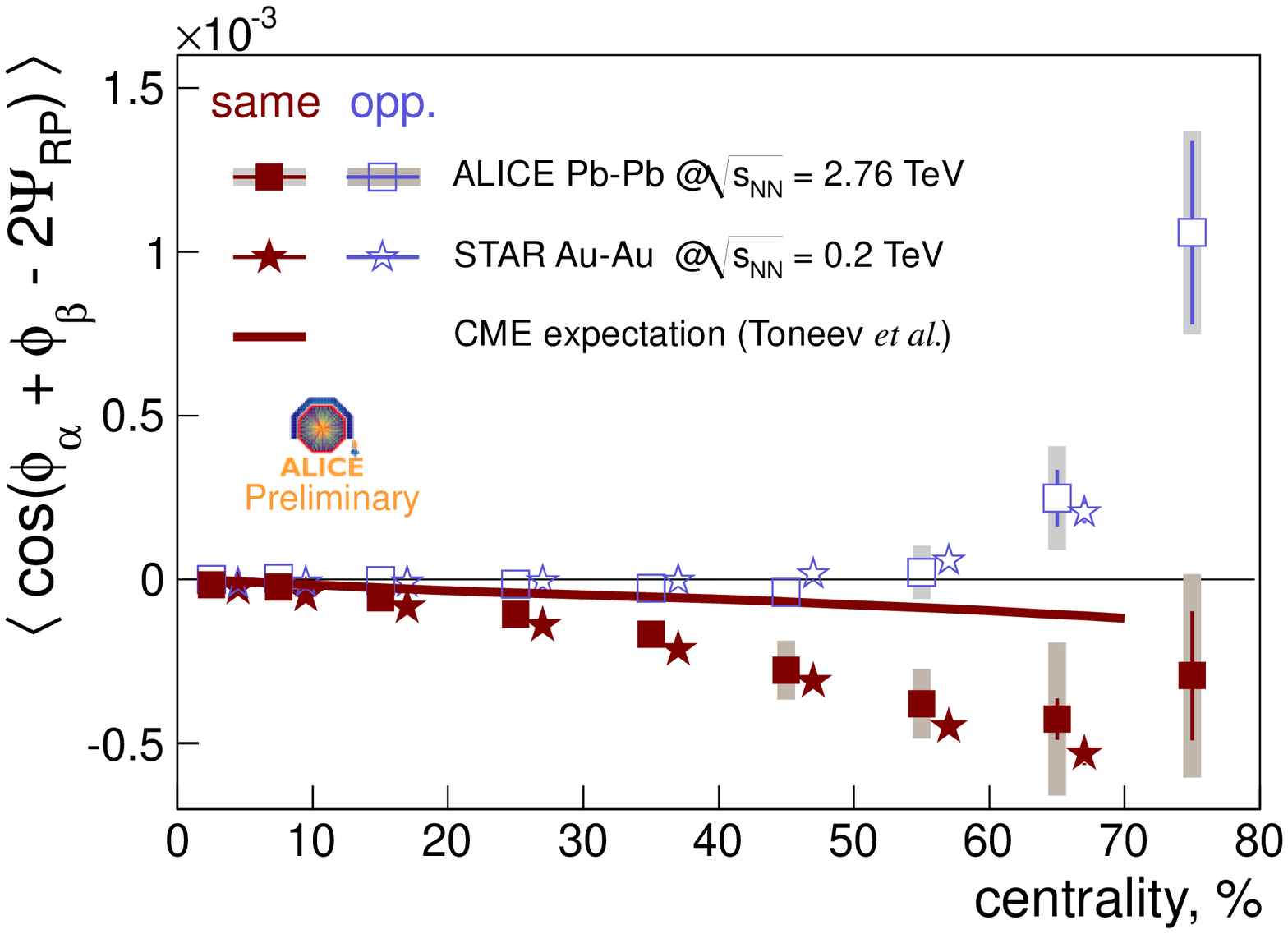}%
\includegraphics[width=0.5\linewidth]{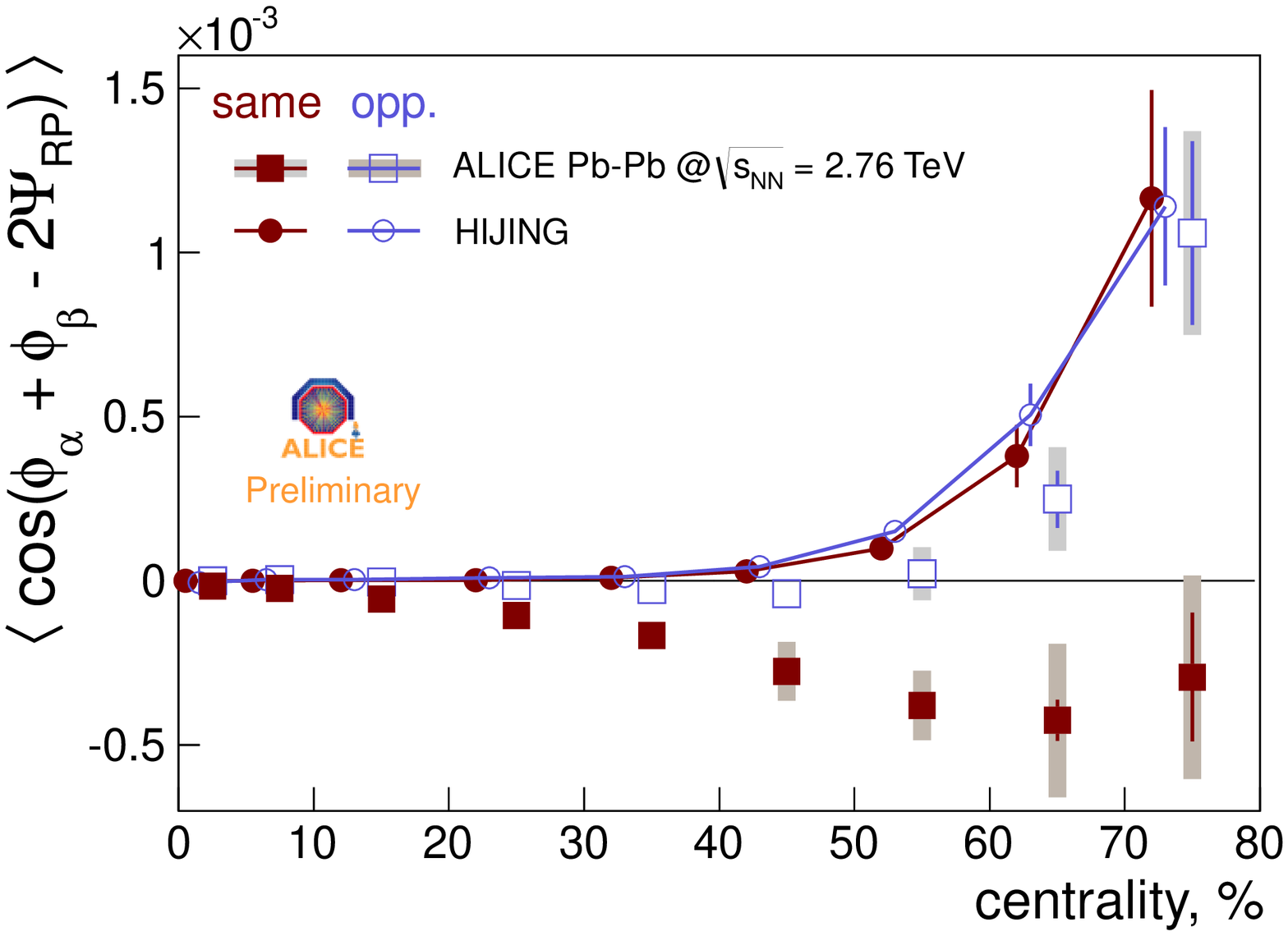}
{\mbox{}\vspace{-0.4cm}
\hspace{3.7cm}\mbox{~} \bf (a)
\hspace{+6.8cm}\mbox{~} \bf (b)}
{\mbox{}\\\vspace{-0.1cm}}
\caption{(color online) 3-particle correlator vs. centrality: (a)
compared to STAR data~\cite{Ref:STAR} and theoretical predictions
which incorporates effects of local parity violation at LHC energies~\cite{Ref:Toneev},
(b) expectations from HIJING~\cite{Ref:Hijing} Monte-Carlo%
% with and without elliptic flow modulations%
.
}
\label{fig:comparisonRP}
\end{figure}

Figure~\ref{fig:comparisonRP}(a) presents the 3-particle correlator (\ref{Eq:3ParticleCorrelator})
measured by the ALICE Collaboration compared to the RHIC data~\cite{Ref:STAR} and correlations
expected for the same charge pairs from effects of local parity violation at LHC energies~\cite{Ref:Toneev}.
Statistical (systematic) uncertainties in Fig.~\ref{fig:comparisonRP}
are shown by the error bars (shaded area) respectively.
In addition to the sources listed in Sec.~\ref{Section:Analysis}, the
dominant contribution to the systematic error comes from the uncertainty in the measurement of the elliptic
flow, $v_2$, which is used to estimate the event plane resolution.
For corrections we used an average of the
elliptic flow measured from 2- and 4-particle cumulants 
\cite{Ref:AliceFlow} which has different sensitivity to the
correlations unrelated to the reaction plane (non-flow) and flow fluctuations.
Half of the difference between v$_2$ measured with 2- and 4-particle
cumulants was assigned as the systematic uncertainty.

Figure~\ref{fig:comparisonRP}(a) reveals a remarkable
similarity between ALICE and STAR measurements performed
at the collision energies $\sqrt{s_{NN}}=0.2$ and 2.76~TeV.
%, what points to little or no energy dependence of the 3-particle correlations.
Both the ALICE and STAR results show a strong correlation between pairs of the same charge,
and, in contrast a weak correlation among oppositely charged pairs.
If the observed charge dependence would be due to only the local parity violation
this behavior can be interpreted as suppression (quenching) by the
medium of the opposite charge sign correlations~\cite{Ref:BField1}.
Another possible explanation for the difference in the magnitude of the correlation
for same and opposite charge sign pairs can be a baseline shift due to charge independent
fluctuations of the directed flow~\cite{Ref:TeaneyYan}.
Together with the ALICE and STAR experimental data, Fig.~\ref{fig:comparisonRP}(a) shows predictions
for the parity violating effects at LHC energies based on a
model~\cite{Ref:Toneev} which incorporates the time evolution of the magnetic field.
The model could not predict the actual magnitude of the
charge correlations, and in the calculations the measured correlations by the
STAR Collaboration were scaled to the LHC energies.
While such predictions underestimate the observed magnitude of the same sign correlations at the LHC
by a factor of five,
it was noticed in \cite{Ref:Zhitnitsky,Ref:BField1}
that the strength of charge correlations
can be the same for RHIC and LHC energies.

Figure~\ref{fig:comparisonRP}(b) shows comparison between the
three particle correlator (\ref{Eq:3ParticleCorrelator})
measured by ALICE and charge correlations extracted from
HIJING~\cite{Ref:Hijing} Monte-Carlo simulations for Pb-Pb collisions
at $\sqrt{s_{NN}} = 2.76$~ TeV.
% The HIJING simulations were performed using the 
% default configuration and modified version which incorporates
% realistic modulations for the elliptic flow.
The correlations in the HIJING simulations show almost no charge dependence,
and significantly increase
for very peripheral collisions,
which is due to correlations not related to the reaction plane
(see for example discussion in \cite{Ref:STAR}). 
Note that in \cite{Ref:Pratt} it is argued that effects of local charge conservation
may be responsible for a significant part of the observed charge
dependence of the correlator~(\ref{Eq:3ParticleCorrelator}).

The differential dependencies of the charge correlations
provides another way to compare various models
which attempt to describe the observed charge separation.
\begin{figure}[ht]
\includegraphics[width=\linewidth]{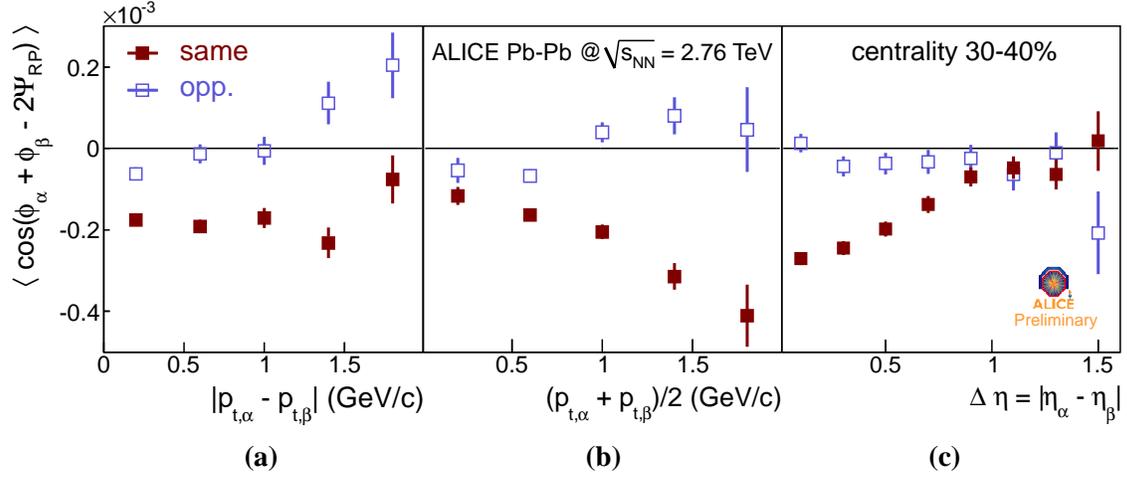}
{\mbox{}\vspace{-0.4cm}
\hspace{3.cm}\mbox{~} \bf (a)
\hspace{+3.8cm}\mbox{~} \bf (b)
\hspace{+3.8cm}\mbox{~} \bf (c)}
{\mbox{}\\\vspace{-0.1cm}}
\caption{
(Color online) 3-particle correlator as a function of the
(a) difference and (b) mean pair transverse momentum,
and (c) the pair separation in pseudorapidity.
}
\label{fig:3pDifferential}
\end{figure}
Figure~\ref{fig:3pDifferential} shows the 3-particle correlator~(\ref{Eq:3ParticleCorrelator})
as a function of the difference and mean pair transverse momentum,
and the pair separation in pseudorapidity for the 30-40$\%$ centrality region.
Compared to pairs of opposite charge, the same charge pairs show a
strong increase in the magnitude with increasing mean
transverse momentum of the pair (Fig.~\ref{fig:3pDifferential}(b)),
with a correlation width of about one unit in pseudorapidity (Fig.~\ref{fig:3pDifferential}(c)).

\section{\label{Section:Summary}Summary}
Two and three particle charge dependent azimuthal correlations have been measured
for Pb-Pb collisions at $\sqrt{s_{NN}} = 2.76$~TeV by ALICE at the LHC.
A clear charge dependence of the correlations is observed.
The three particle correlations are very similar to those reported by the
STAR Collaboration for Au-Au collisions at $\sqrt{s_{NN}} = 0.2$~TeV.
At the same time, the two particle correlations differ strongly from those observed at RHIC.
Together, the two and  three particle correlations
measured at the LHC may provide strong experimental constraints on the possible mechanism of
parity conserving and parity odd effects in heavy-ion collisions.

\section*{\label{Section:acknowledgments}Acknowledgment}
This work was supported by the Helmholtz Alliance Program of the Helmholtz Association,
contract HA216/EMMI "Extremes of Density and Temperature: Cosmic Matter in the Laboratory".

\end{document}